\documentclass[%
 reprint, 
superscriptaddress,
 amsmath,amssymb,
 aps, pre,
]{revtex4-2}

\usepackage{graphicx}
\usepackage{dcolumn}
\usepackage{bm}
\usepackage{xcolor}

\usepackage{tikz}
\usepackage{tikz-3dplot}
\usepackage{physics}
\definecolor{pivgreen}{RGB}{190,255,120}

\begin{document}


\title{\textbf{Experimental evaluation of Kolmogorov's first similarity hypothesis for shear-dependent viscosity fluids}}

\author{L. Hergenreder}
\affiliation{Departamento de Ingeniería Hidráulica y Ambiental, Pontificia Universidad Católica de Chile, Av. Vicuña Mackenna 4860, 7820436 Santiago, Chile}

\author{S. Simo\"ens}
\affiliation{Laboratoire de Mécanique des Fluides et d’Acoustique, UMR CNRS 5509, Université Claude Bernard Lyon 1, INSA de Lyon, Ecole Centrale de Lyon, 20 avenue Albert Einstein, F-69621, Villeurbanne, France}

\author{W. Brevis}%
 \email{Contact author: wbrevis@uc.cl}
\affiliation{Departamento de Ingeniería Hidráulica y Ambiental, Pontificia Universidad Católica de Chile, Av. Vicuña Mackenna 4860, 7820436 Santiago, Chile}

\date{\today}

\begin{abstract}
We experimentally investigate the applicability of Kolmogorov's first similarity hypothesis to shear-dependent viscosity fluids. We assess whether dissipative range statistics exhibit a self-similar behavior in nearly homogeneous and isotropic turbulence, using stereo particle image velocimetry in aqueous Carbopol solutions across two concentrations and a range of forcing conditions. We observe that, when normalized by the mean viscosity and mean dissipation rate, second- and third-order structure functions collapse across the different experimental conditions and exhibit functional forms compatible with classical Newtonian results. These observations are in general agreement with the recent numerical results of Rosti [J. Fluid Mech. \textbf{1012}, R5 (2025)], who proposed that Kolmogorov's similarity hypotheses can be extended to shear-thinning fluids by considering dissipation scales based on the mean viscosity, the mean dissipation rate, and the flow index.
\end{abstract}

\maketitle


\section{\label{sec:introduction}Introduction}
Turbulence is a state of chaotic fluid motion involving many degrees of freedom, making a statistical treatment the only viable approach \citep{monin2007statistical,mccomb2014homogeneous}. In this context, obtaining a universal description of the flow movement, applicable to a wide range of conditions, has been a longstanding objective of the turbulence community. This knowledge has fundamental implications for many processes, such as mixing \citep{shralman2000scalar}, the propagation of sound and light waves, and combustion \citep{sreenivasan1997phenomenology}. In addition, the formulation of turbulence models, such as large-eddy simulation (LES) or Reynolds-averaged Navier-Stokes (RANS) models, relies on the realistic parametrization of the small scales \citep{sreenivasan1997phenomenology}.

A fundamental contribution in this regard was made by \citet{kolmogorov1991local}, who postulated a series of hypotheses leading to the idea of the universal behavior of the small scales of turbulence. Kolmogorov stated that the directionality of the large scales, corresponding to the scales of energy injection, is progressively lost during the process of energy transfer towards smaller scales. Consequently, below a certain size, the small-scale motions become statistically isotropic, defining the so-called universal equilibrium range. In the first similarity hypothesis, Kolmogorov postulated that the statistics of motion in the universal equilibrium range are uniquely determined by the viscosity $\nu_0$ and the mean rate of turbulent kinetic energy dissipation $\langle\varepsilon\rangle$. As a result, when normalized by these two quantities, the statistics of velocity take a universal form, independent of the specific flow configuration. Dimensional analysis of $\nu_0$ and $\langle \varepsilon \rangle$ yields the smallest scales of motion, or Kolmogorov scales, for length $\eta = (\nu_0^3/\langle \varepsilon \rangle)^{1/4}$, time $\tau_\eta = (\nu_0 / \langle \varepsilon \rangle)^{1/2}$, and velocity $u_\eta = (\nu_0 \langle \varepsilon \rangle)^{1/4}$. Scales comparable to $\eta$, for which viscous effects are significant, define the dissipation range. In contrast, the larger scales within the universal equilibrium range form the inertial subrange, whose statistics, according to the second similarity hypothesis, are determined uniquely by $\langle \varepsilon \rangle$, independently of $\nu_0$.

This phenomenology, however, considers a constant Newtonian viscosity, while many fluids in nature and industry exhibit more complex, non-Newtonian behaviors. A common example is given by fluids exhibiting yield stress and shear-thinning effects, which generally act together and result in a viscosity that is determined by the local flow conditions. These fluids are frequently found under turbulent conditions, such as during the transport of mining tailings, domestic slurries, concrete, drilling muds and food products, as well as in geophysical contexts like particle-laden, pyroclastic, and lava flows \citep{thota2018rheological,chhabra2011,Frigaard2019}. In these materials, shear thinning refers to a reduction in viscosity with increasing shear rate, while yield stress behavior implies that the material behaves as a solid until the applied stress exceeds a threshold value, the yield stress $\tau_y$, after which it flows \citep{chhabra2011}. While these effects can coexist with thixotropy (a dependence on the prior history of deformation of the material) or elasticity (the capacity to partially recover the imposed deformation) \citep{chhabra2011}, we address only purely viscous, shear-dependent effects in this work.

Under turbulent conditions, the viscosity in shear-dependent viscosity fluids is not a constant material property, but a quantity with spatiotemporal variability. %
Consequently, Kolmogorov's first similarity hypothesis, which refers to the smallest scales of motion, is not trivially applicable. It is not clear, in principle, how to define an appropriate viscosity scale which determines the smallest scales of motion, nor whether such a scale would allow velocity statistics in the dissipation range to be expressed in a self-similar form. Recently, \citet{rosti2025effect} addressed this issue by performing direct numerical simulations (DNS) of forced HIT flows of shear-thinning fluids, with viscosity following a power-law dependence on the shear rate. The author suggested that Kolmogorov's similarity hypotheses can be extended to shear-thinning fluids by considering the mean viscosity $\langle \nu \rangle$ instead of the Newtonian viscosity $\nu_0$, together with the flow index $n$, the exponent of the power law relating viscosity and shear rate. Considering these quantities, the first extended similarity hypothesis states that the statistics of the small scales of motion are uniquely determined by the mean viscosity $\langle \nu \rangle$, the mean dissipation rate $\langle \varepsilon\rangle$, and the flow index $n$, while the second similarity hypothesis states that, for the intermediate range of scales, the statistics are uniquely determined by $\langle \varepsilon\rangle$.

Although the extension of the Kolmogorov hypotheses allowed \citet{rosti2025effect} to express the energy spectra and structure functions of HIT flows of purely shear-thinning fluids in a self-similar form, and to obtain inertial range scalings equivalent to those of Newtonian fluids, it remains unclear whether this extension holds more generally for shear-dependent-viscosity fluids. When yield stress is analyzed in isolation (in the absence of shear-thinning) through DNS of HIT flows, rheologically dependent scaling is observed for spectra and structure functions \citep{abdelgawad_scaling_2023}, an effect that is not expected within the extended hypotheses. Additionally, studies addressing the combined effect of both behaviors, corresponding to experimental \citep{mitishita2021fully} and numerical \citep{narayanan2024turbulent} works in pipe and channel flows, report energy spectra with scaling ranges that differ from those obtained in DNS of isotropic turbulence when each effect is considered separately. Whether these differences arise from the flow configuration or from the interaction between the two non-Newtonian behaviors remains unclear, highlighting the incomplete understanding of small-scale behavior in turbulent flows of shear-dependent viscosity fluids.

In this context, the objective of this article is to experimentally investigate the applicability of Kolmogorov's first similarity hypothesis to fluids with shear-dependent viscosity. We focus specifically on the first hypothesis, rather than the second, because it concerns the dissipation range, where the fluctuating viscosity is expected to play a determining role. To this end, we employ stereo particle image velocimetry (SPIV) to evaluate homogeneous and isotropic flows of aqueous solutions of Carbopol. We assess the applicability of the first similarity hypothesis by examining second- and third-order structure functions, and explore whether analytical expressions can be given to describe their behavior.

\section{\label{sec:methods}Materials and methods}

\subsection{\label{subsec:device}HIT chamber}
\begin{table*}
\caption{\label{tab:table1}Summary of experimental conditions: propeller frequency $f_p$, Reynolds number of the installation $Re_I = f_p D^2/\langle \tilde{\nu} \rangle_{ctr,t}$, velocity scale $\mathcal{U}=\langle u_{\mathrm{rms}} \rangle_{ctr}$, longitudinal integral length scale $L_L$, mean kinetic energy dissipation rate $\langle \tilde{\varepsilon} \rangle_{ctr,t}$, mean kinematic viscosity $\langle \tilde{\nu} \rangle_{ctr,t}$, Kolmogorov length scale $\eta$ and velocity scale $u_\eta$, and Taylor microscale Reynolds number $Re_\lambda = \mathcal{U}\lambda/\langle \tilde{\nu} \rangle_{ctr,t}$, where $\lambda = (15 \langle \tilde{\nu} \rangle_{ctr,t} \, \mathcal{U}^2 / \langle \tilde{\varepsilon} \rangle_{ctr,t})^{1/2}$ is the Taylor microscale. See Sec.~\ref{subsec:averages} for definitions of the averages employed, and Sec.~\ref{subsec:scales} for the estimation of the dissipative scales.}
\begin{ruledtabular}
\begin{tabular}{lccccccccc}
Fluid & $f_p$ [Hz] & $Re_I $ & $ \mathcal{U}$ [mm/s] & $L_L$ [mm] & $\langle \tilde{\varepsilon} \rangle_{ctr,t}$  [mm\textsuperscript{2}/s\textsuperscript{3}] & $\langle \tilde{\nu} \rangle _{ctr,t}$ [mm\textsuperscript{2}/s] & $\eta$ [mm] & $u_\eta$ [mm/s] & $Re_\lambda $\\
\hline
C=0.04\% & 3.00 & 148 & 20.12 & 83.40 & 482.84 & 202.51 & 11.45 & 17.68 & 5.0\\
C=0.04\% & 4.00 & 225 & 24.86 & 77.23 & 776.97 & 177.98 & 9.23 & 19.28 & 6.4\\
C=0.04\% & 5.02 & 336 & 33.82 & 66.77 & 1614.45 & 149.60 & 6.57 & 22.17 & 9.0\\
C=0.04\% & 5.99 & 475 & 45.53 & 64.54 & 3283.35 & 125.97 & 4.97 & 25.36 & 12.5\\
C=0.04\% & 7.17 & 688 & 61.19 & 56.00 & 7052.56 & 104.21 & 3.56 & 29.28 & 16.9 \\
C=0.05\% & 4.00 & 35 & 13.15 & 151.63 & 472.04 & 1140.68 & 42.11 & 27.09 & 0.9\\
C=0.05\% & 5.02 & 70 & 28.05 & 113.50 & 1815.25 & 719.69 & 21.29 & 33.82 & 2.7\\
C=0.05\% & 5.99 & 94 & 32.35 & 110.79 & 2437.77 & 638.70 & 18.08 & 35.32 & 3.2\\
C=0.05\% & 7.17 & 128 & 38.44 & 104.83 & 3461.91 & 557.52 & 14.96 & 37.27 & 4.1\\
\end{tabular}
\end{ruledtabular}
\end{table*}

\begin{figure}
\includegraphics[width=.45\textwidth]{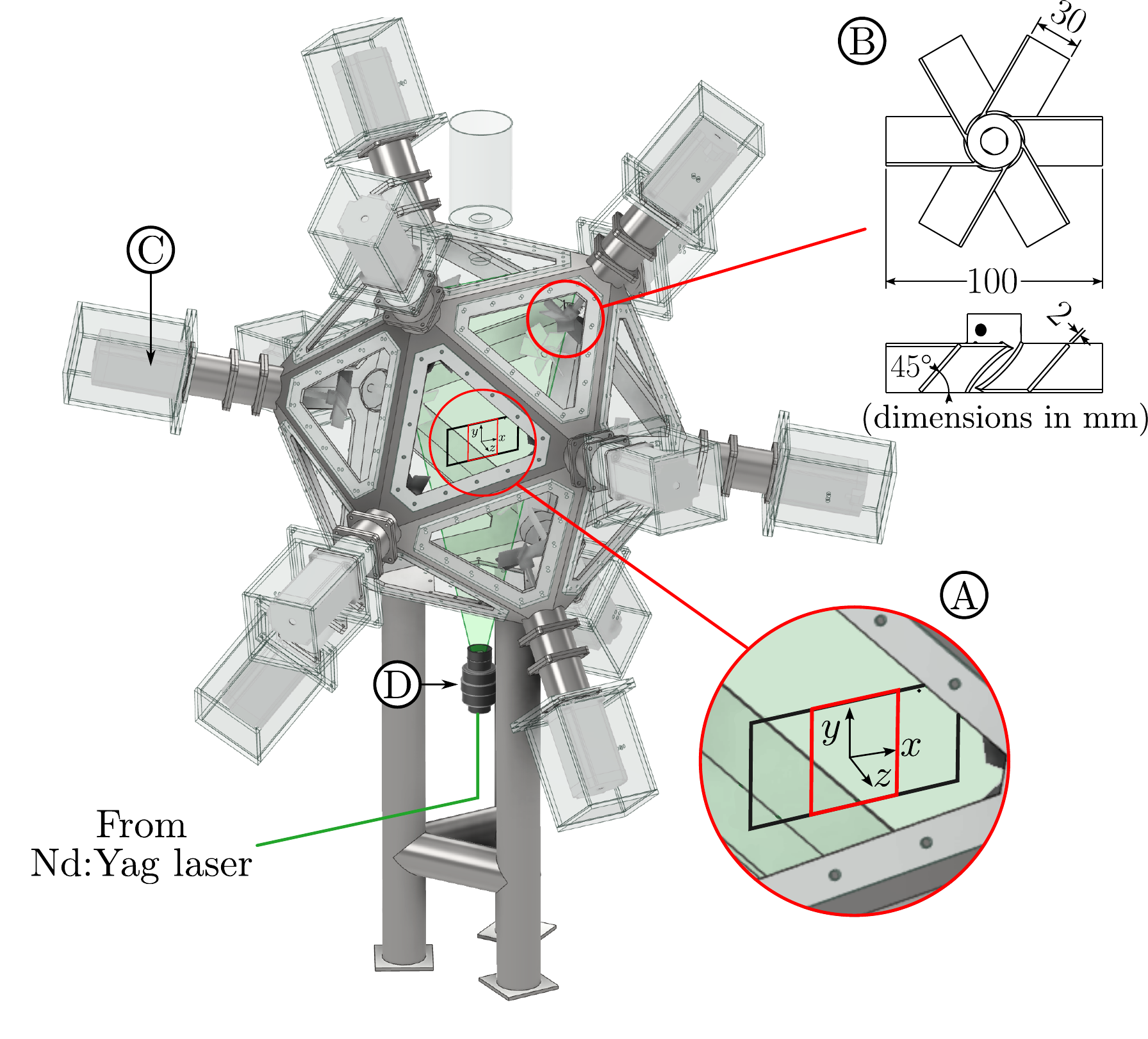}
\caption{\label{fig:device} Experimental device. Field of view (black rectangle) and central HIT region (red square) with coordinate axis (A); propeller design (B); electric motor (C); laser sheet optics (D).}
\end{figure}

To generate the flow conditions for this study, we employ a propeller-driven HIT chamber (Figure~\ref{fig:device}). This device is based on the design of \citet{zimmermann2010lagrangian} and similar to that of \citet{fiabane2012clustering} and \citet{wang2025lagrangian}. It consists of an icosahedral chamber that encloses 140 liters of the test fluid. Twelve center-pointing propellers are installed at the chamber vertices and, driven by electric motors, are employed to establish the flow conditions. The propellers have a diameter $D=10$ cm and a pitched blade geometry with six blades, as shown in Fig.~\ref{fig:device} B. In this work, different flow conditions are generated by operating all propellers simultaneously at equal and constant frequencies $f_p$, between 3 and 7.17 Hz. To characterize the different experimental conditions, we consider the Reynolds number of the installation, $Re_I = f_p D^2 / \langle \tilde{\nu} \rangle_{ctr,t}$, where $\langle \tilde{\nu} \rangle_{ctr,t}$ is the estimated viscosity averaged over time and the homogeneous region (see Sec.~\ref{subsec:scales} and Table~\ref{tab:table1}). 

This experimental configuration approximates a spherical forcing, which is known to produce, for Newtonian fluids, nearly homogeneous and isotropic turbulence with a reduced mean flow around the chamber center, as observed in previous studies \citep{ghazi_nezami_laboratory_2023}. For the non-Newtonian fluids employed in this study, we have performed an extensive characterization of the overall flow conditions generated by this device in Ref.~\citep{hergenreder2026experimental}. By evaluating velocity and viscosity statistics, we observed that the flow can be considered nearly homogeneous and isotropic within the region $|x|/D , |y|/D\leq0.3$, where the origin of coordinates is set at the chamber center (Figure~\ref{fig:device} A). A summary of the main flow scales characterizing the experimental conditions is presented in Table~\ref{tab:table1}.

\subsection{\label{subsec:velocimetry}Velocity measurements}
In order to study the dynamics of the generated flows, we measure the three velocity components in a vertical plane passing through the central point of the chamber by means of stereo particle image velocimetry (SPIV). The measurement plane is determined by a laser sheet entering the chamber through the bottom window, which is generated by a pulsed Nd:YAG laser along with a $-25$~mm cylindrical lens and a collimator, producing a sheet approximately 2~mm thick in the measurement region. Test fluids are seeded with hollow glass spheres with a mean diameter $d_p$ of 10 to 13 $\mu m$ and a density $\rho_p$ of 1.5~g/cm\textsuperscript{3}, yielding Stokes numbers below 10\textsuperscript{-6} and ensuring they behave as passive tracers \citep{adrian2011particle}. The seeded flow is imaged by two 12 MP cameras equipped with 85~mm lenses, and Scheimpflug mounts to properly focus the measurement plane. These cameras image a region around the chamber center with lines of sight perpendicular to two of the windows, avoiding the need for water prisms.  This results in viewing angles of $\alpha _{xz} \approx 37 ^{\circ}$ and $\alpha _{yz} \approx -23^{\circ} $, measured in the horizontal ($xz$) and vertical planes ($yz$) respectively.

The two-dimensional three-component (2D3C) velocity fields are obtained using an in-house implementation of the method of \citet{soloff_distortion_1997}. To this end, 2D2C velocity fields are first obtained for each camera, by processing particle images with PIVlab \citep{thielicke_particle_2021}. First, images are enhanced by subtracting the mean intensity, applying a contrast-limited adaptive histogram equalization, and performing low-pass filtering. Planar velocity fields are then obtained for each camera by applying a PIV analysis with two passes ($64 \times 64$~px and $32 \times 32$~px, with 50\% overlap), resulting in less than 3\% of outlier vectors for all cases. Subsequently, the 2D3C velocity field is reconstructed from the previously obtained 2D2C fields by means of polynomial mapping functions as described in Ref.~\citep{soloff_distortion_1997}. Mapping functions are obtained by capturing images of a calibration target at $z$ positions in the range $-2\leq z \leq 2$~mm in steps of $0.5$~mm. For this process, the target is carefully positioned by means of a dedicated system mounted onto the HIT chamber structure, consisting of a rotational and a linear stage that provide a resolution of $0.1$ degrees in the orientation of the $z$ axis, and $0.1$ mm in the $z$ position of the calibration plane. By determining the coordinates of the calibration points with subpixel accuracy by fitting skewed Gaussian kernels to the recorded intensity distributions \citep{Raffel2018}, a polynomial mapping function of fifth order in $x$ and $y$ and third order in $z$ was obtained, with root mean square values of the reprojection error below $0.5$ px. The resulting field of view (FOV) covers a region of $141 \times 65 $ mm, and the vector spacing is $\Delta x = 0.69$~mm. For every experimental condition, 7250 image pairs are collected at a sampling frequency of 1 Hz in groups of 250, which was sufficient to converge the main statistics analyzed in this work.

\subsection{\label{subsec:fluids}Test fluids}
We employ aqueous Carbopol Ultrez 10 (Lubrizol Corporation) solutions at concentrations of 0.04 and 0.05 wt.\% as test fluids. Aqueous Carbopol solutions are commonly used as model fluids since they exhibit both shear-thinning and yield stress behaviors with only minor thixotropic and elastic effects \citep{Varges2019rheological}. This leaves viscous effects dominating the rheology of the fluid, making these solutions good approximations of purely viscous, shear-dependent viscosity fluids. Additionally, their transparency allows implementing optical velocimetry techniques. The test fluids are prepared by dispersing the Carbopol powder into distilled water, followed by neutralization using a 20\% NaOH solution \citep{hergenreder2026experimental}. To prevent lump formation, the Carbopol powder is sieved before incorporation \citep{Varges2019rheological}, and the neutralizing solution is added drop by drop during mixing, which promotes polymer swelling and gel formation.

\begin{figure}
\includegraphics[width=.45\textwidth]{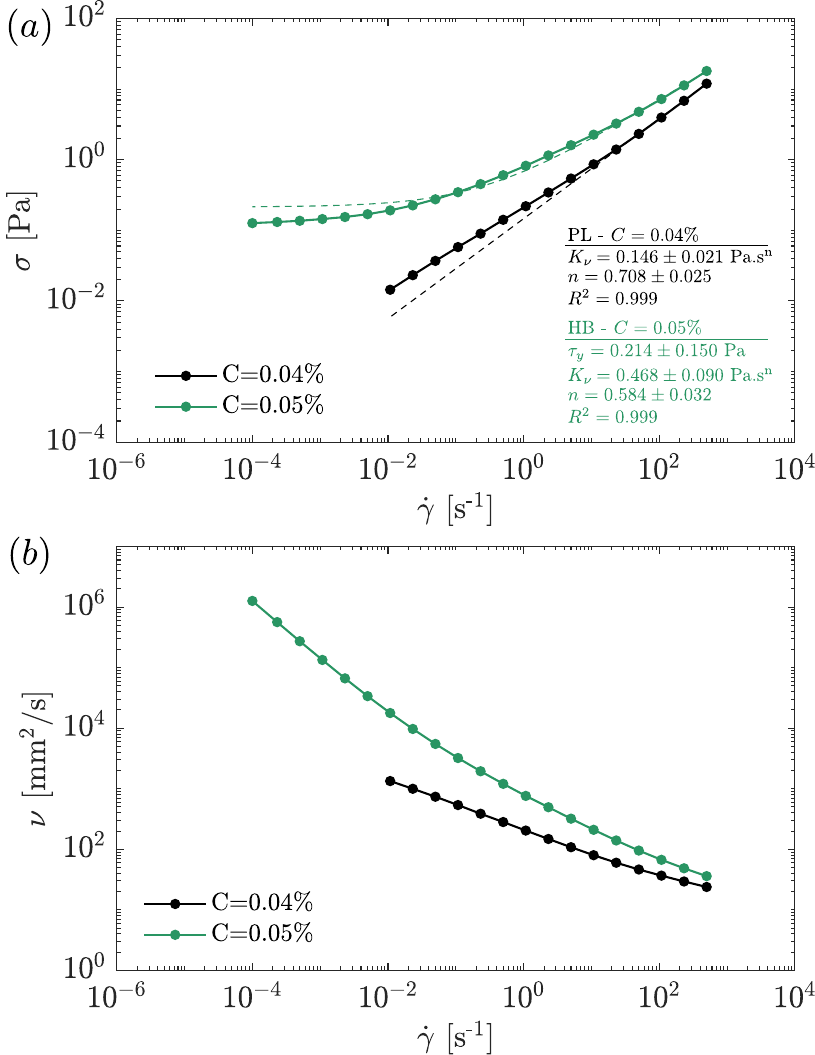}
\caption{\label{fig:flow_curves} Flow curves for the aqueous solutions of Carbopol at concentrations of 0.04\% (black) and 0.05\% (green). (a) Shear stress as a function of the shear rate, with dashed lines showing fits to the Herschel-Bulkley model (Eq.~\ref{eq:HB_model}). (b) Kinematic viscosity as a function of the shear rate.}
\end{figure}

To characterize the viscous behavior of the test fluids, flow curves were measured with a TA Instruments ARES-G2 rheometer for the C=0.05\% fluid and a DHR-3 rheometer for the C=0.04\% fluid, using samples extracted after the velocimetry runs. A grooved Couette geometry was used in both cases to prevent wall slip. Shear stress $\sigma$ was measured as a function of shear rate $\dot{\gamma}$ over $\dot{\gamma} = [0.001,500]$ s\textsuperscript{-1} for the C=0.05\% fluid and $\dot{\gamma} = [0.01,500]$ s\textsuperscript{-1} for the C=0.04\% fluid. For the latter, the lower bound was determined by the lowest measurable torque of the instrument, since the low viscosity of the solution leads to torque values below the sensor resolution for shear rates lower than $10^{-2}$ s\textsuperscript{-1}. The measured range, however, covered the shear rates observed in the center of the chamber during the velocimetry measurements, according to shear rate values estimated from our velocimetry measurements (see Sec.~\ref{subsec:scales}). Each curve was measured from high to low shear rates, which decreases the duration of the tests and avoid startup effects observed when measuring from low to high $\dot \gamma$ \citep{Varges2019rheological}. At every point, the torque was measured in steady state conditions, following the same criterion as Ref.~\citep{Varges2019rheological}. To provide a reference for the viscous behavior of the fluids, we fit the data to the Herschel-Bulkley model,
\begin{equation}\label{eq:HB_model}
\begin{aligned}
\dot{\gamma} &= 0                   && \text{if }\;\sigma < \tau_y\,,  \\
\sigma         &= \tau_y + K_{\mu} \dot{\gamma}^n   && \text{otherwise}
\end{aligned}
\end{equation}

\noindent where $\tau_y$, $K_{\mu}$, and $n$ are the yield stress, consistency index, and flow index, respectively. The fits yield $\tau_y= 0.214$~Pa, $n=0.584$, and $K_{\mu}=0.468$ Pa$\cdot$s$^{n}$ for the more concentrated (C=0.05\%) fluid, and $\tau_y=0.055$~Pa, $K_{\mu}=0.136$ Pa$\cdot$s$^{n}$, $n=0.719$ for the C=0.04\% fluid, as shown in Fig.~\ref{fig:flow_curves}(a). Overall, these measurements show that the aqueous solutions of Carbopol employed in this study are shear-thinning fluids with limited yield stress effects.

\subsection{\label{subsec:averages}Averages and two-point statistics}
In this work, we consider a Cartesian coordinate system $(x,y,z)$ with its origin at the geometric center of the chamber and axes oriented as shown in Fig.~\ref{fig:device}A. From the two-dimensional, three-component velocity field measured by SPIV, $V_i(\bm{x},t)$, the mean and fluctuating components are defined as $U_i(\bm{x}) = \langle V_i (\bm{x},t)\rangle_t$ and $u_i(\bm{x},t) = V_i(\bm{x},t) - U_i(\bm{x})$, respectively, where $\langle \cdot \rangle_t$ denotes temporal averaging.

As mentioned in Sec.~\ref{subsec:device}, in a previous study \citep{hergenreder2026experimental} we characterized the flow properties of aqueous solutions of Carbopol in the HIT chamber, observing that within $|x|/D \leq 0.3$ and $|y|/D \leq 0.3$, the flow can be considered nearly homogeneous and isotropic with a negligible mean flow. We take advantage of this property for the computation of a series of statistics below, by considering their spatial average over this central region, which we denote by $\langle \cdot \rangle_{ctr}$.

\begin{figure}[t]
\centering
    \tdplotsetmaincoords{70}{120}
    \begin{tikzpicture}[
        tdplot_main_coords,
        axis/.style={ -latex},
        vec/.style={thick, red!75!black, -latex},
        dashedline/.style={dashed, gray},
        arc/.style={gray, -latex}
    ]
        \def\vx{2.5}
        \def\vy{2.5}
        \def\vz{3.0}

        \pgfmathsetmacro{\Rlen}{sqrt(\vx*\vx + \vy*\vy + \vz*\vz)}
        \pgfmathsetmacro{\xzLen}{sqrt(\vx*\vx + \vz*\vz)}
        
        \pgfmathsetmacro{\phiAngle}{atan(\vx / \vz)}

        \pgfmathsetmacro{\thetaAngle}{acos(\vy / \Rlen)}

        \pgfmathsetmacro{\vxNorm}{\vx / \xzLen}
        \pgfmathsetmacro{\vzNorm}{\vz / \xzLen}

        \coordinate (O) at (0,0,0);
        
        \coordinate (X_axis) at (0, 3.0, 0);
        \coordinate (Y_axis) at (0, 0, 3.5);
        \coordinate (Z_axis) at (4.0, 0, 0);
        
        \coordinate (R) at (\vx, \vy, \vz);
        
        \coordinate (P_xz) at (\vx, 0, \vz);
        \coordinate (P_z)  at (\vx, 0, 0);
        \coordinate (P_y)  at (0, 0, \vz);
        
        \coordinate (P_vis_xy) at (0, \vy, \vz);
        \coordinate (P_vis_xz) at (\vx, \vy, 0);
        \coordinate (P_vis_x)  at (0, \vy, 0);

        \filldraw[fill=pivgreen, draw=pivgreen!60!black, fill opacity=0.7] 
            (0, -1.2, 0) -- (0, 2.25, 0) -- (0, 2.25, 3.25) -- (0, -1.2, 3.25) -- cycle;
        \filldraw[fill=pivgreen, draw=pivgreen!60!black, fill opacity=0.7] 
            (0, -1.2, 0) -- (0, 2.25, 0) -- (0, 2.25, -1.25) -- (0, -1.2, -1.25) -- cycle;
            
        \path (0, 0.5, 3.25) -- (0, 2.25, 3.25) 
            node[midway, above, sloped, text=pivgreen!60!black, font=\scriptsize, align=right] 
            {PIV plane};
        
        \draw[dashedline] (P_y) -- (P_xz) -- (P_z);
        \draw[dashedline] (O) -- (P_xz);
        \draw[dashedline] (P_xz) -- (R);

        \draw[dashedline] (R) -- (P_vis_xz);
        \draw[dashedline] (P_vis_x) -- (P_vis_xz) -- (P_z);

        \draw[dashedline] (R) -- (P_vis_xy);
        \draw[dashedline] (P_vis_x) -- (P_vis_xy) -- (P_y);

        \draw[axis] (O) -- (X_axis) node[right] {$x$};
        \draw[axis] (O) -- (Y_axis) node[above] {$y$};
        \draw[axis] (O) -- (Z_axis) node[below left] {$z$};
        
        \node[below, xshift=4pt, yshift=-2pt] at (O) {$\bm{X}_0$};

        \draw[vec] (O) -- (R) node[anchor=west, yshift=-4pt, xshift=-2pt] {$\bm{r}/2$};

        \tdplotsetthetaplanecoords{0}
        \tdplotdrawarc[tdplot_rotated_coords, arc]{(O)}{1.6}{0}{\phiAngle}{anchor=south east, yshift = -1pt, xshift=5pt}{$\phi$}
        
        \def\arcRadius{2.0}
        \draw[arc] plot[domain=0:\thetaAngle, samples=30, smooth, variable=\t] 
            ({\arcRadius * sin(\t) * \vxNorm}, 
             {\arcRadius * cos(\t)}, 
             {\arcRadius * sin(\t) * \vzNorm});
        
        \def\lblRadius{2.15}
        \pgfmathsetmacro{\midTheta}{\thetaAngle / 1.5}
        \node[right, xshift=2pt] at 
            ({\lblRadius * sin(\midTheta) * \vxNorm}, 
             {\lblRadius * cos(\midTheta)}, 
             {\lblRadius * sin(\midTheta) * \vzNorm}) {\textcolor{gray}{$\theta$}};
             
    \end{tikzpicture}
\caption{Spherical coordinate system $(r,\theta,\phi)$ centered at a point $\bm{X}_0$. The polar angle $\theta$ is measured between the separation vector $\bm{r}$ and the $x$-axis. The SPIV measurement plane (green) corresponds to the $xy$ plane, or equivalently, $\phi=0$.}
\label{fig:spherical_coords}
\end{figure}
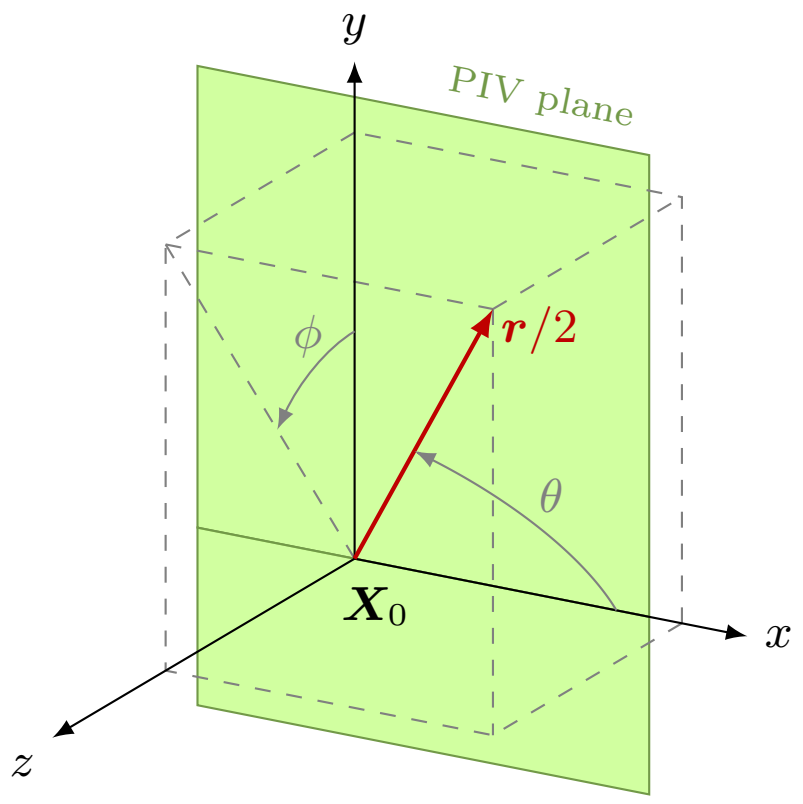

To compute two-point statistics in this study, owing to the nearly spherical symmetry of the forcing in the experimental device, we define velocity increments relative to a central point $\bm{X}_0$. Around $\bm{X}_0$, we consider pairs of points $\bm{x}^+$ and $\bm{x}^-$ separated by a displacement vector $\bm{r}$, such that $\bm{x}^+ = \bm{X}_0 +\frac{1}{2} \bm{r}$ and $\bm{x}^- = \bm{X}_0 -\frac{1}{2} \bm{r}$. Velocity increments between these points are then given by $\delta u_i (\bm{X}_0,\bm{r},t) = u_i(\bm{x}^+,t) - u_i(\bm{x}^-,t)$. For convenience, we define a spherical coordinate system $(r,\theta,\phi)$ centered at $\bm{X}_0$, where $r =\lVert \bm{r} \rVert$ is the scalar separation, $\theta$ is the polar angle between $\bm{r}$ and the $x$ axis, and $\phi$ is the azimuthal angle. Note that $\phi=0$ corresponds to our measurement plane $xy$ (Fig.~\ref{fig:spherical_coords}).

To investigate the behavior of two-point statistics defined from velocity increments $\delta u_i (\bm{X}_0,\bm{r},t)$, we consider a spherical average over all orientations of $\bm{r}$ at fixed scalar separation $r$. For an arbitrary scalar function $f(\bm{r}) = f(r,\theta, \phi)$, we denote the spherical averaging operator as \citep{lawson_unsteady_2022}

\begin{align}
    \langle f \rangle_\circ (r) &= \frac{1}{4\pi} \int_0 ^{2\pi} \int_0 ^{\pi} f(r,\theta,\phi) \sin \theta \, \mathrm{d} \theta \, \mathrm{d} \phi \nonumber \\
     &\approx \frac{1}{2} \int_0 ^\pi f(r,\theta,\phi=0) \sin \theta \, \mathrm{d}\theta \label{eq:spherical_avg}
\end{align}

\noindent where we have assumed symmetry around the $x$ axis to approximate the spherical average from our measurements of $f(\bm{r})$ in the $\phi=0$ ($xy$) plane. Below we use the averages defined here to estimate the Kolmogorov scales and structure functions from our experimental data.

\section{\label{sec:results}Results}
\subsection{\label{subsec:scales}Dissipative scales}
For a purely viscous, shear-dependent viscosity fluid, the instantaneous kinematic viscosity is completely determined by the instantaneous shear rate, $\nu(\bm{x},t) = \varphi ( \dot{\gamma}(\bm{x},t))$, where $\varphi(\cdot)$ captures the non-Newtonian behaviors, such as shear-thinning or yield stress. The shear rate is given by $\dot \gamma = (2 S^2)^{1/2}$, where $S^2$ is the squared strain rate tensor $S_{ij}$, defined as $S_{ij} = 1/2(\partial V_i/\partial x_j + \partial V_j/\partial x_i)$. Additionally, for these variable viscosity fluids, the instantaneous dissipation rate results, $\varepsilon (\bm{x},t) = 2 \nu (\bm{x},t)S^2(\bm{x},t)$.

The direct determination of $\nu(\bm{x},t)$ and $\varepsilon(\bm{x},t)$ therefore requires access to the full velocity gradient tensor $\partial V_i / \partial x_j$ to evaluate $S^2$. Since our SPIV measurements are restricted to a two-dimensional plane, we employ the following surrogate expression to estimate the squared strain rate \citep{sharp_piv_2001},

\begin{eqnarray}
    \widetilde{S^2} (\bm{x},t) = \biggl( \frac{\partial V_1}{\partial x_1} \biggr) ^2 + \biggl( \frac{\partial V_2}{\partial x_2} \biggr) ^2 + \biggl( \frac{\partial V_2}{\partial x_1} \biggr) ^2 \nonumber \\
    + \frac{1}{2} \biggl( \frac{\partial V_1}{\partial x_2} \biggr) ^2 + \frac{1}{2}\biggl( \frac{\partial V_1}{\partial x_2} + \frac{\partial V_2}{\partial x_1} \biggr) ^2 \label{eq:squared_strain}
\end{eqnarray}

To minimize the effect of noise, spatial gradients are estimated from velocity fields recomputed at half the original resolution. We previously evaluated the surrogate expression (\ref{eq:squared_strain}) by comparing water flow measurements in our setup with DNS of Newtonian HIT flows at comparable Reynolds numbers, finding good agreement in the probability density function of $S^2$. Further details on the computation of these statistics and the validation procedure are provided in Ref.~\citep{hergenreder2026experimental}.

Using Eq.~(\ref{eq:squared_strain}), we obtain the local, instantaneous viscosity and dissipation fields. First, we compute the shear rate as $\widetilde{\dot \gamma}(\bm{x},t) = (2 \widetilde{S^2})^{1/2}$. To this end, we first estimate the local viscosity $\tilde{\nu}(\bm{x},t)$ by employing the measured flow curve (Figure~\ref{fig:flow_curves}(b)) as a lookup table, with $\widetilde{\dot \gamma}(\bm{x},t)$ as the query variable. Finally, we obtain the instantaneous dissipation from the estimated viscosity and squared strain rate as $\tilde{\varepsilon}(\bm{x},t) = 2 \tilde{\nu}(\bm{x},t) \widetilde{ S^2}(\bm{x},t)$.

\begin{figure}
\includegraphics[width=.47\textwidth]{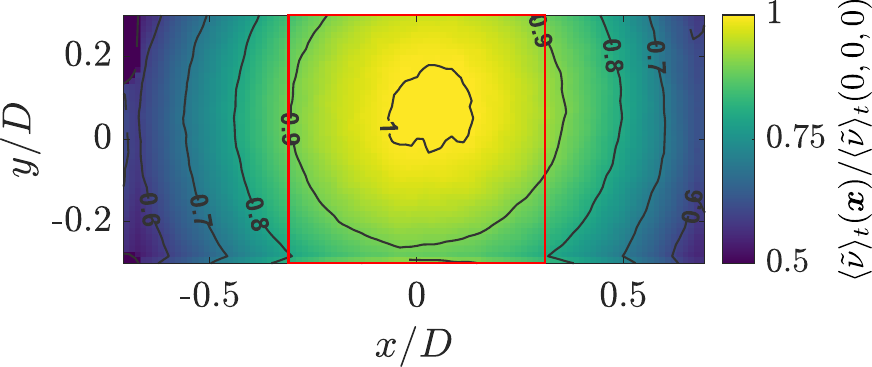}
\caption{\label{fig:viscosity_homogeneity} Viscosity homogeneity: time-averaged viscosity field normalized by its value at the chamber center $(x,y,z)=(0,0,0)$, for a representative case (C=0.05\% fluid forced at $f_p=7.17$ Hz). The red square denotes the region of spatial averaging $|x|/D,|y|/D \leq 0.3$ (see Sec.~\ref{subsec:averages}).}
\end{figure}

We then compute the Kolmogorov scales for all flow conditions. According to \citet{rosti2025effect}, the classical similarity hypotheses can be extended to shear-thinning fluids by considering the mean viscosity and mean dissipation rate as the parameters determining the statistics of the smallest flow scales. Following this, we estimate the mean viscosity $\langle \tilde{\nu} \rangle_{ctr,t}$ and dissipation rate $\langle \tilde{\varepsilon} \rangle_{ctr,t}$ by averaging these quantities over time and the central region of our measurement domain $|x/D| \leq 0.3 $, $|y/D| \leq 0.3 $. The resulting values are presented in Table~\ref{tab:table1}. It is worth noting that the time-averaged viscosity field is highly homogeneous across the central region for both fluids and all flow conditions. The local variations relative to the chamber center, $\langle \tilde{\nu} \rangle_t(\bm{x}) / \langle \tilde{\nu} \rangle_t(0,0,0)$, remain within 10\% (see Fig.~\ref{fig:viscosity_homogeneity}). Furthermore, twice the spatial standard deviation relative to the mean value $2\sigma_{ \langle \tilde{\nu} \rangle_t } / \langle \langle \tilde{\nu} \rangle_t \rangle_{ctr}$ is below 0.1 \citep{hergenreder2026experimental}. 

Considering the viscosity and dissipation rate averaged over space and time, $\langle \tilde{\nu} \rangle_{ctr,t}$ and $\langle \tilde{\varepsilon} \rangle_{ctr,t}$, the length and velocity scales of the small-scale structures can be estimated as

\begin{align}
    \eta &= \left( \left\langle \tilde{\nu} \right\rangle_{ctr,t}^3 / \left\langle \tilde{\varepsilon} \right\rangle_{ctr,t} \right)^{1/4} \label{eq:eta}\\
    u_\eta &= \left( \left\langle \tilde{\nu} \right\rangle_{ctr,t}  \left\langle \tilde{\varepsilon} \right\rangle_{ctr,t} \right)^{1/4} \label{eq:u_eta}
\end{align}

\noindent where $\eta$ and $u_\eta$ are the Kolmogorov length and velocity scales, respectively. The values obtained for each experimental condition are presented in Table~\ref{tab:table1}.

\subsection{\label{subsec:S2}Second-order structure functions}
To assess the impact of the shear-dependent viscosity on the dynamics of the small scales, we first examine the longitudinal second-order structure function $S_2(r)$. Using the velocity increments and the spherical average introduced in Sec.~\ref{subsec:averages}, $S_2$ is defined as

 \begin{align}
     S_2(r) = \left\langle D_{ij}  \frac{r_i r_j}{r^2} \right\rangle_{\circ} \label{eq:S2}
 \end{align}

\noindent where $D_{ij}(\bm{r}) = \langle \delta u_i \delta u_j \rangle_{ctr,t}$ is the second-order structure function tensor. Both temporal and spatial averages are applied here for consistency with the third-order structure function (Eq.~\ref{eq:S3}), where spatial averaging is employed to improve statistical convergence.

\begin{figure*}
\includegraphics[width=.85\textwidth]{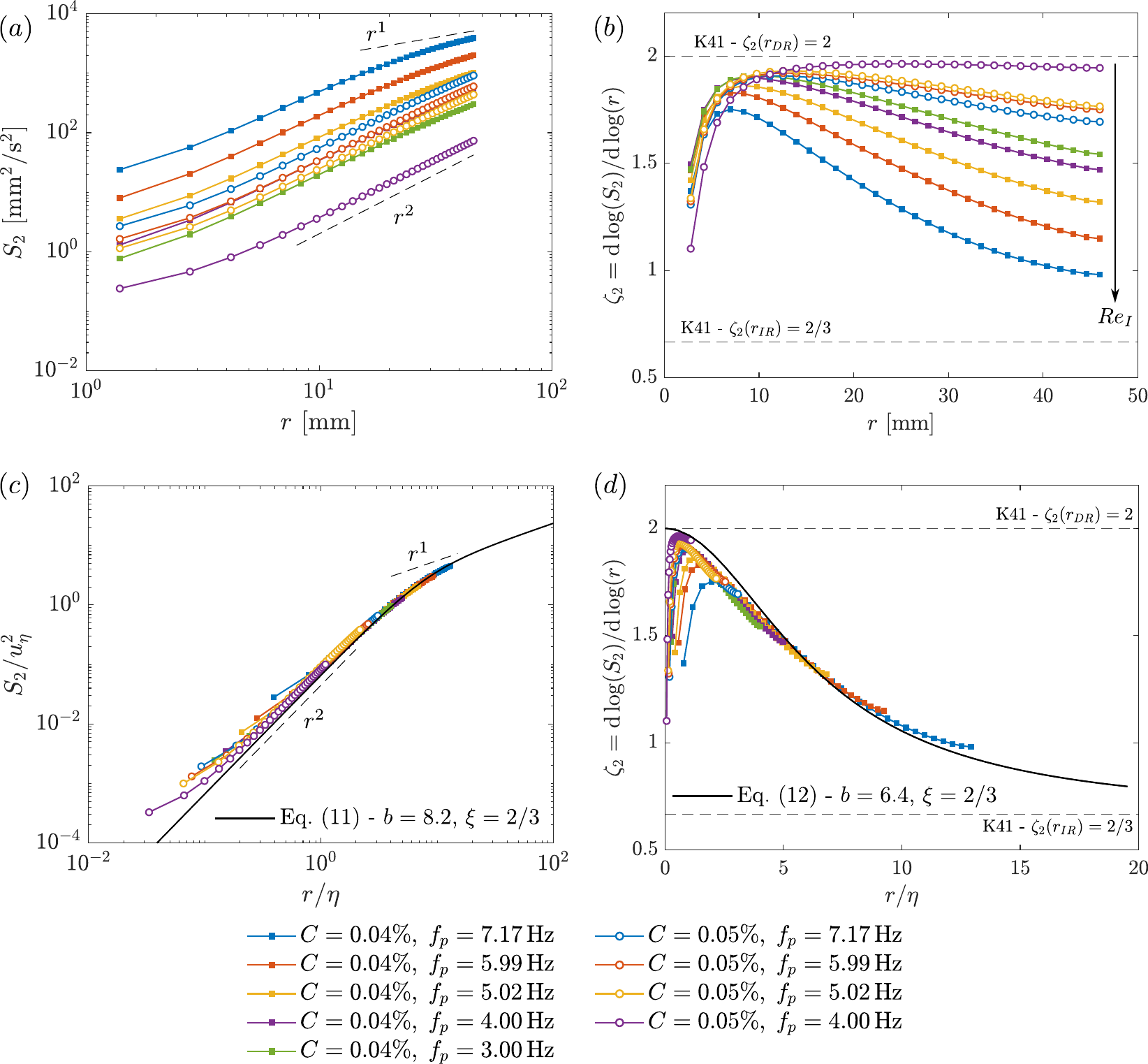}
\caption{\label{fig:S2} (a) Second-order structure functions $S_2$ versus separation distance $r$ for all experimental conditions. (b) Local slope $\zeta_2(r)$ versus separation distance $r$ for all experimental conditions. Upper and lower dashed lines correspond to the Kolmogorov predictions for the dissipative and inertial ranges, respectively. (c) Second-order structure functions $S_2$ compensated by dissipative scales, plotted versus normalized separation distance $r/\eta$. The solid line corresponds to the Batchelor parametrization (Eq.~\ref{eq:batchelor_selfsimilar}) with $b=8.2$ and $\xi=2/3$. (d) Local slope $\zeta_2$ versus normalized separation distance $r/\eta$. Upper and lower dashed lines correspond to the Kolmogorov predictions for the dissipative and inertial ranges, respectively. The solid line corresponds to the scaling exponents resulting from the Batchelor parametrization (Eq.~\ref{eq:batchelor_exponents}) with $b=6.4$ and $\xi=2/3$.}
\end{figure*}

We compute the second-order longitudinal structure functions for both test fluids and all forcing conditions (Fig.~\ref{fig:S2}(a)). It can be observed that all cases fall within the curves corresponding to the least concentrated fluid forced at the highest propeller frequency (blue squares) and the most concentrated fluid at the lowest propeller frequency (violet hollow circles), which correspond to the highest and lowest $Re_I$, respectively. For these two limiting cases, the largest measured separations approximately follow power-law scalings of $r^1$ and $r^2$ for the highest and lowest $Re_I$, respectively. To evaluate these scaling ranges more precisely, we compute the local scaling exponent,

\begin{align}
    \zeta_2(r) \equiv \frac{\mathrm{d} \log (S_2)}{\mathrm{d} \log (r)} \label{eq:local_exponent_S2}
\end{align}

\noindent which corresponds to the local slope of $S_2$ in a logarithmic plot, such that $S_2 \sim r^{\zeta_2(r)}$ (Fig.~\ref{fig:S2}(b)). These exponents provide a clearer distinction between the different experimental conditions. For separations $r \gtrsim 15$~mm, $\zeta_2(r)$ decreases systematically as $Re_I$ increases. At the largest measured separations, the curves present approximately constant values of $\zeta_2(r)$, between $1$ and $2$.

For high-Reynolds-number Newtonian flows, a result of the similarity hypotheses \citep{kolmogorov1991local} is that, in the inertial and dissipative ranges the second-order structure functions should scale as $S_2 \sim r^{2/3}$ and $S_2 \sim r^{2}$, respectively. To contextualize our Carbopol results, we indicate these values as dashed lines in Fig.~\ref{fig:S2}(b). We observe that the effect of increasing $Re_I$ seems to be to progressively transition from the dissipative to the inertial predictions for the largest measured separations, $r \gtrapprox 40 \, \text{mm}$.

Given the resemblance of the structure function scalings between the Carbopol cases and the Newtonian predictions, we now turn to evaluate whether our non-Newtonian results also agree with the first self-similarity hypothesis extended by \citet{rosti2025effect}. This implies,

\begin{align}
    S_2/u_\eta ^2 = F \left( r/\eta \right) \label{eq:S2_similarity}
\end{align}

\noindent where $F(\cdot)$ is a scaling function, and $u_\eta$ and $\eta$ are the Kolmogorov velocity and length scales defined in Sec.~\ref{subsec:scales}.

We test the applicability of the self-similarity hypothesis to our experimental data by plotting the compensated second-order structure functions following Eq.~(\ref{eq:S2_similarity}) in Fig.~\ref{fig:S2}(c). We observe that the results corresponding to both fluids and all forcing conditions collapse onto a single curve, confirming the existence of the self-similar function $F(\cdot)$. Moreover, we observe that the $r^2$ range holds for separations around the Kolmogorov scale $r \approx \eta$, and that $r^1$ corresponds to scales $r \approx 10\eta$. We also present the local scaling exponent as a function of the normalized separation $r/\eta$ in Fig.~\ref{fig:S2}(d). In line with the second-order structure function results, $\zeta_2(r)$ also collapses onto a single curve for the different experimental conditions when normalized.

These results support the idea of $S_2$ presenting a universal, self-similar form across different flow conditions, which implies the existence of the scaling function $F(\cdot)$ in Eq.~(\ref{eq:S2_similarity}). To investigate the extent to which this function is compatible with that observed for Newtonian fluids, we consider the Batchelor parametrization of the second-order structure function \citep{batchelor1951pressure},

\begin{align}
    S_2(r) = s_{xx}^2 \frac{r^2}{[1 + (r/r_d)^2]^{1-\xi/2}} \label{eq:batchelor_normal}
\end{align}

\noindent where $r_d$ is the crossover scale between the inertial and dissipative ranges, $s_{xx}^2 = \langle (\partial u_1 / \partial x_1)^2 \rangle$ is the mean squared longitudinal velocity gradient, and $\xi = 2/3$ is the inertial range scaling exponent. Although Eq.~(\ref{eq:batchelor_normal}) can be regarded as an interpolation formula rather than a result derived from first principles, it has been extensively used to characterize second-order structure functions \citep[e.g. ][]{lohse1995bottleneck,zhang2021experimental}, making it a suitable reference to evaluate whether complex fluids recover the main properties of Newtonian phenomenology.

To compare the analytical expression with our experimental results, we rewrite Eq.~(\ref{eq:batchelor_normal}) in normalized form. As a first step, we consider isotropic conditions and approximate the squared velocity gradient in terms of the mean viscosity and mean dissipation rate as $s_{xx}^2 \approx \langle \varepsilon \rangle / 15 \langle \nu \rangle$. This leads to,

\begin{align}
    \frac{S_2 (r)}{u_\eta ^2} = \frac{1}{15} \frac{(r/\eta)^2}{\left[ 1 + (1/b)^2 (r/\eta)^2 \right]^{1-\xi/2}} \label{eq:batchelor_selfsimilar}
\end{align}

\noindent where $b=r_d /\eta$ is the crossover scale relative to the dissipative length scale. In the dissipative range ($r \ll \eta$), Eq.~(\ref{eq:batchelor_selfsimilar}) recovers $S_2/u_\eta ^2 \sim (r/\eta)^2$, whereas in the inertial subrange ($r \gg \eta$) it yields $S_2/u_\eta^2 \sim (r/\eta)^{2/3}$.

We then set $\xi=2/3$ and fit the crossover scale in Eq.~(\ref{eq:batchelor_selfsimilar}) to our experimental data, obtaining $b = 8.2 \pm 1.30$. Using this value, we plot in Figure~\ref{fig:S2}(c) the Batchelor parametrization and observe that it presents a good agreement with the normalized experimental curves, consequently being a good representation of $F(\cdot)$ in Eq.~(\ref{eq:S2_similarity}). Furthermore, the $r^1$ scaling lies in the transition between the dissipation and inertial ranges, and is compatible with a transition between both asymptotic cases, rather than an anomalous scaling.

To evaluate whether this agreement extends to the scaling exponents, we also compare the experimental values with those predicted by the Batchelor parametrization. Differentiating Eq.~(\ref{eq:batchelor_selfsimilar}) yields the analytical expression for the local scaling exponent,

\begin{equation}
    \zeta_2 (r/\eta) = 2 - \frac{(2 - \xi)(1/b)^2(r/\eta)^2}{1+(1/b)^2(r/\eta)^2} \label{eq:batchelor_exponents}
\end{equation}

Fitting Eq.~(\ref{eq:batchelor_exponents}) with $\xi=2/3$ to the experimentally obtained exponents yields $b=6.4 \pm 0.13$. Employing these values to plot Eq.~(\ref{eq:batchelor_exponents}) shows an excellent agreement with the experimental data (Fig.~\ref{fig:S2}(d)). We note the obtained value of $b$ is below that obtained from the normalized second-order structure function $S_2(r)/u_\eta^2$. We attribute this difference to the fact that $S_2(r)/u_\eta^2$ is fitted in log-log coordinates while $\zeta_2$ is fitted in linear coordinates, which results in a different sensitivity of the fit across the measured scale range.

\subsection{\label{subsec:S3}Third-order structure functions}
We now turn to the evaluation of the third-order longitudinal structure function $S_3(r)$. As described for $S_2$, we consider the velocity increments and spherical averaging introduced in Sec.~\ref{subsec:averages} to obtain

 \begin{align}
     S_3(r) = \left\langle D_{ijk}  \frac{r_i r_j r_k}{r^3} \right\rangle_{\circ} \label{eq:S3}
 \end{align}

 \noindent where $D_{ijk}(\bm{r}) = \langle \delta u_i \delta u_j \delta u_k\rangle_{ctr,t}$ is the third-order structure function tensor. As noted in Sec.~\ref{subsec:S2}, spatial averaging is applied here in addition to time averaging to improve the convergence of this odd-order statistic.

\begin{figure*}
\includegraphics[width=.8\textwidth]{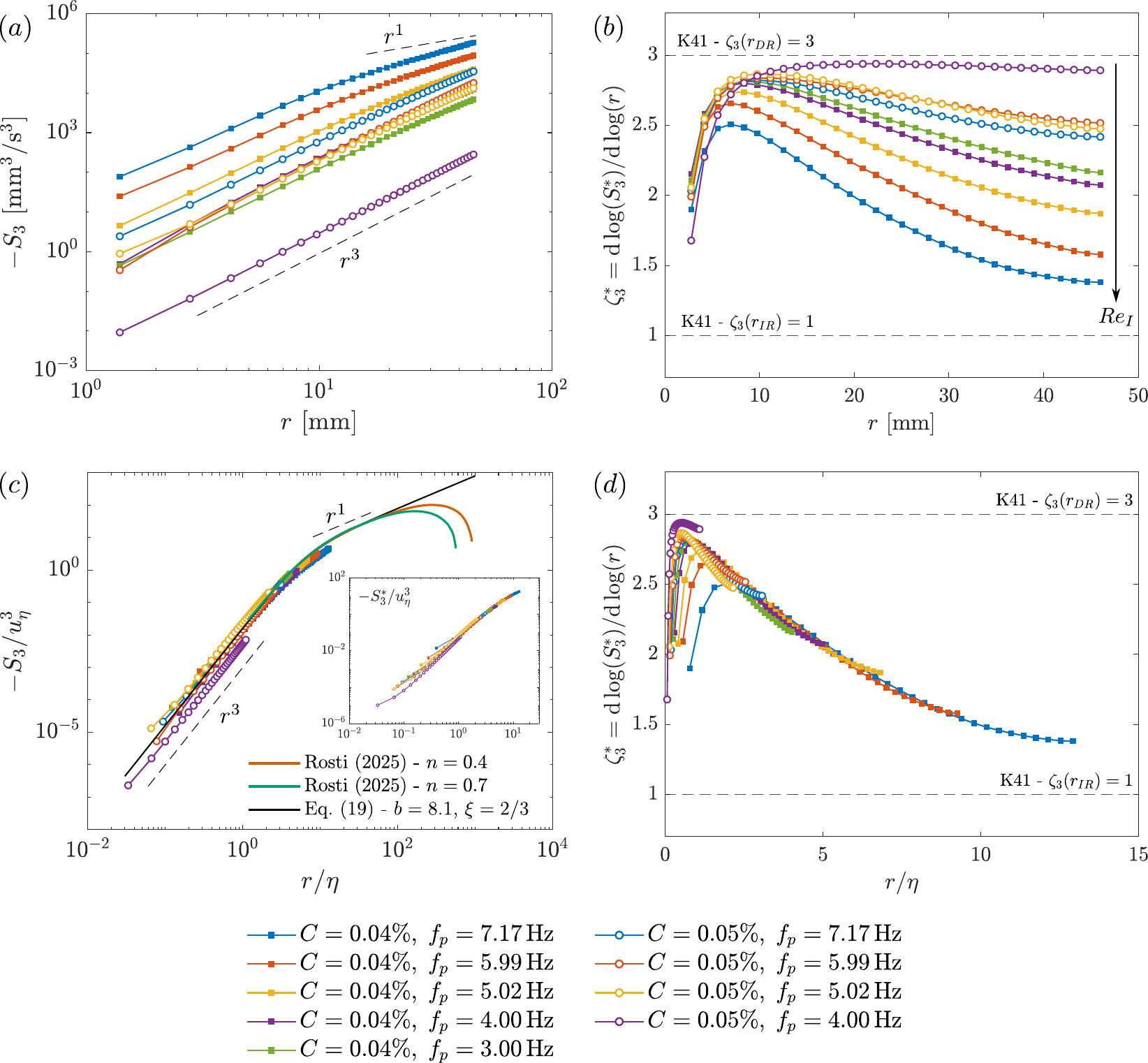}
\caption{\label{fig:S3} (a) Third-order structure functions $S_3$ versus separation distance $r$ for all experimental conditions. (b) Local slope $\zeta^*_3(r)$ versus separation distance $r$. Upper and lower dashed lines correspond to the Kolmogorov predictions for the dissipative and inertial ranges, respectively. (c) Third-order structure functions compensated by dissipative scales versus normalized separation distance $r/\eta$, along with the analytical expression Eq.~(\ref{eq:S3_kolmogorov_normalized}) with $b=8.1$ and $\xi=2/3$ (solid black line). Orange and green lines show the DNS results of \citet{rosti2025effect} for HIT at $Re_\lambda \approx 320$ in Carreau fluids with flow indices of 0.4 and 0.6, respectively. The inset in (c) shows the collapse of $-S_3^*/u_\eta^3$ versus $r/\eta$ for all experimental conditions. (d) Local slope $\zeta_3^*$ versus normalized separation distance $r/\eta$. Upper and lower dashed lines correspond to the Kolmogorov predictions for the dissipative and inertial ranges, respectively. In panels (b) and (d), we considered structure functions computed from the absolute value of velocity increments, $S_3^* (r)$, to improve statistical convergence.}
\end{figure*}

Computing $S_3(r)$ for each flow condition results in the curves presented in Fig.~\ref{fig:S3}(a). The curve corresponding to the lowest Reynolds number follows an $r^3$ scaling, while the one with the highest $Re_I$ approaches $r^1$. To evaluate these scaling regimes more closely, we compute the local scaling exponent $\zeta_3^*$, defined as

\begin{align}
    \zeta_3^* \equiv \frac{\mathrm{d} \log (S_3^*)}{\mathrm{d} \log( r)} \label{eq:local_exponent_S3}
\end{align}

\noindent where we consider the third-order structure function computed from the absolute value of velocity increments, $S_3^*$, to improve statistical convergence. A comparison between the curves across the different flow conditions shows a similar picture to that of the second-order structure functions: for the largest measured separations ($r \gtrsim 40 $~mm) the case with the lowest Reynolds number presents a slope compatible with the dissipative range scaling, $S_3 \sim r^3$, while increasing $Re_I$ systematically decreases the local slope toward the inertial range scaling predicted by the classical Kolmogorov phenomenology, $S_3 \sim r^1$.

Following the approach used for $S_2$, we test the applicability of the extended first similarity hypothesis to the third-order statistics by normalizing the curves with the dissipative scales defined in Sec.~\ref{subsec:scales} \citep{rosti2025effect}. According to this hypothesis, $S_3$ should exhibit a universal, self-similar form given by

\begin{align}
    S_3/u_\eta^3 = G(r/\eta) \label{eq:S3_similarity}
\end{align}

\noindent where $G(\cdot)$ is a scaling function. When normalized by the Kolmogorov scales computed from the mean viscosity and dissipation, the experimental data for the different flow conditions collapse onto a single curve (Fig.~\ref{fig:S3}(c)), with the only exception of the lowest Reynolds number case (C = 0.05\% forced at $f_p = 4$ Hz), which falls slightly below the rest of the curves. However, if the normalization is applied to $S_3^*$, which is based on the absolute value of the velocity increments, all cases successfully collapse (see the inset of Fig.~\ref{fig:S3}(c)). Consequently, we attribute the lack of collapse in the original $S_3$ curve to a limited statistical convergence in the lowest $Re_I$ case. As observed for second order statistics, plotting the local scaling exponent $\zeta_3^*$ against the normalized separation $r/\eta$ also shows a remarkable collapse for all flow conditions, as shown in Fig.~\ref{fig:S3}(d), further supporting the idea of a universal behavior across flow conditions.

Given that the second- and third-order structure functions for the Carbopol solutions present a self-similar behavior and scalings compatible with Newtonian turbulence, it is natural to investigate whether the Kolmogorov equation, which relates both quantities, is also satisfied. For locally isotropic Newtonian turbulence, the Kolmogorov equation reads \citep{kolmogorov1991dissipation},

\begin{equation}
    S_3(r) = -\frac{4}{5} \langle \varepsilon \rangle r + 6 \nu_0 \frac{\partial S_2 (r)}{\partial r} \label{eq:kolmogorov_newtonian}
\end{equation}
\noindent where $\nu_0$ is the Newtonian kinematic viscosity.

To obtain an analogous expression for fluids with shear-dependent viscosity, we assume that the mean viscosity is homogeneous over the analyzed domain. As commented in Sec.~\ref{subsec:scales}, this is a reasonable approximation for the statistics generated around the geometrical center of our experimental device (see Fig.~\ref{fig:viscosity_homogeneity}). Then, by taking the Reynolds decomposition of the viscosity, and following the standard derivation \citep{mccomb2014homogeneous}, we obtain

\begin{equation}
    S_3(r) = -\frac{4}{5} \langle \varepsilon \rangle r + 6 \langle \nu \rangle \frac{\partial S_2 (r)}{\partial r} + N(r) \label{eq:kolmogorov_non-newtonian}
\end{equation}

\noindent where the impact of viscosity fluctuations $\nu_{\text{fluc}} = \nu - \langle \nu \rangle$ is captured by the term $N(r) = (1/r^3) \int_0 ^r s^2 \mathcal{V}_{ii}(s) \mathrm{d}s$ with $\mathcal{V}_{ij} = \partial / \partial r_k (\langle 2\nu_{\text{fluc}}^+ s_{ik}^+ u_j^- - 2 \nu_{\text{fluc}}^- s_{jk}^- u_i^+ \rangle )$. Here, $s_{ij}$ is the fluctuating strain rate, and superscripts $+$ and $-$ denote quantities evaluated at positions $\bm{x}^+$ and $\bm{x}^-$, respectively (see Sec.~\ref{subsec:averages}).

Comparing the Newtonian and shear-dependent viscosity expressions (Eqs.~\ref{eq:kolmogorov_newtonian} and \ref{eq:kolmogorov_non-newtonian}) shows that the variable viscosity introduces an extra term involving viscosity fluctuations, while the remaining part of the expression preserves their classical Newtonian structure, provided the Newtonian viscosity $\nu_0$ is replaced by the mean value $\langle \nu \rangle$.

To investigate the form of the self-similar scaling function $G(\cdot)$ in Eq.~(\ref{eq:S3_similarity}), we consider the extended Kolmogorov equation (\ref{eq:kolmogorov_non-newtonian}) to obtain an analytical expression for $S_3(r)$. Assuming a negligible contribution from the viscosity fluctuation term $N(r)$, we substitute the Batchelor parametrization (Eq.~\ref{eq:batchelor_normal}) into the mean viscous transport term $6\langle \nu \rangle \partial S_2 / \partial r$, which leads to

\begin{equation}
    S_3(r) = \underbrace{-\frac{4}{5} \langle \varepsilon \rangle r}_{I_I} + \underbrace{\frac{4}{5} \langle \varepsilon \rangle r \ \frac{1 + (\xi/2) (r/r_d)^2}{[1 + (r/r_d)^2]^{2-\xi/2}} }_{I_V}\label{eq:S3_kolmogorov_regular}
\end{equation}

\noindent where $I_I$ and $I_V$ are the inertial and mean viscous terms, respectively.

We can finally obtain an expression for the scaling function $G(\cdot)$ in Eq.~(\ref{eq:S3_similarity}) by dividing both sides of Eq.~(\ref{eq:S3_kolmogorov_regular}) by $\langle \varepsilon \rangle \eta = u_\eta ^3$ and recalling that $b=r_d/\eta$, which leads to

\begin{equation}
    \frac{S_3(r)}{u_\eta^3} = -\frac{4}{5} \left( \frac{r}{\eta} \right) + \frac{4}{5} \left( \frac{r}{\eta} \right)  \frac{1 + (\xi/2) (r/\eta)^2 (1/b)^2}{[1 + (r/\eta)^2 (1/b)^2]^{2-\xi/2}}\label{eq:S3_kolmogorov_normalized}
\end{equation}

Equations~(\ref{eq:S3_kolmogorov_regular}) and (\ref{eq:S3_kolmogorov_normalized}) present the contribution of the inertial term, $I_I=-4/5\langle \varepsilon \rangle r$, and the mean viscous term $I_V$ to the third-order structure function. For large scales, $r \gg \eta$, the viscous term vanishes, resulting in the well-known four-fifths law, $S_3 \approx -4/5 \langle \varepsilon \rangle r$. On the other hand, in the dissipation range ($r \ll \eta$), the mean viscous term $I_V$ tends to $4/5 \langle \varepsilon \rangle r$, counteracting the inertial term $I_I$ such that $S_3 \to 0$.

We first evaluate the ability of the analytical prediction (Eq.~\ref{eq:S3_kolmogorov_normalized}) to describe the normalized third-order structure functions obtained experimentally. To this end, we set $\xi=2/3$ and fit the crossover length scale $b$ to the experimental $S_3$ data (excluding the lowest $Re_I$ case), which results in $b=8.1 \pm 0.34$. This is in excellent agreement with the value of $b=8.2$ obtained by fitting the Batchelor parametrization to $S_2$. As shown in Fig.~\ref{fig:S3}(c), the analytical expression (black solid line) correctly describes the experimental data. Furthermore, we also include the normalized third-order structure functions obtained numerically by \citet{rosti2025effect} from DNS of HIT flows of Carreau fluids at $Re_\lambda \approx 320$ with flow indices $n = 0.4$ and $0.7$ (orange and green solid lines). The numerical data collapse well with our experimental results over the overlapping range ($1 \lessapprox r/\eta \lessapprox 15$), with only minor deviations near $r/\eta \approx 10$. In addition, the DNS results show a remarkable agreement with the analytical prediction of Eq.~(\ref{eq:S3_kolmogorov_normalized}), displaying the expected linear scaling for separations within the inertial range ($r/\eta \gtrsim 10$).

\begin{figure}
\includegraphics[width=.48\textwidth]{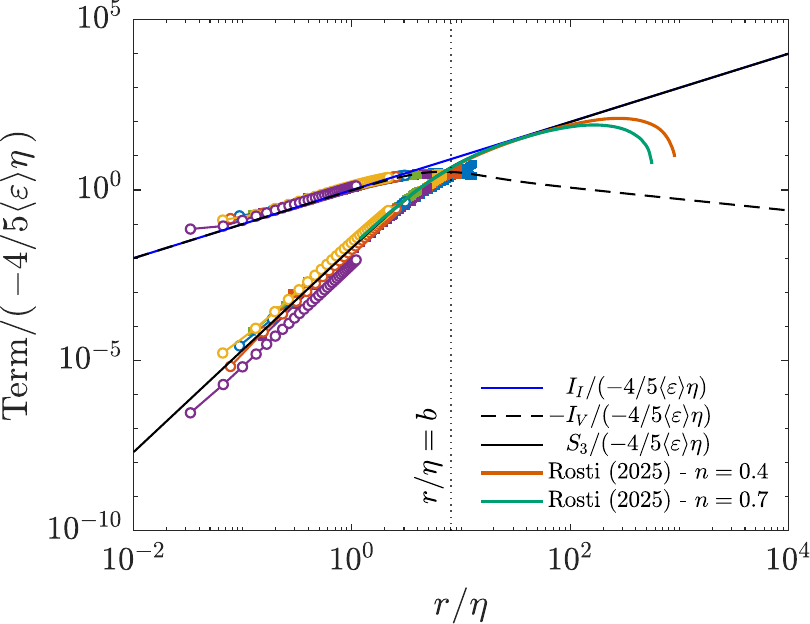}
\caption{\label{fig:kolmogorov_equation} Contribution of the terms in the modified Kolmogorov equation (\ref{eq:S3_kolmogorov_regular}), normalized by $(-4/5)\langle \varepsilon \rangle \eta$. Orange and green lines show the DNS results of \citet{rosti2025effect} for HIT at $Re_\lambda \approx 320$ in Carreau fluids with flow indices of 0.4 and 0.6, respectively. The dotted line denotes the crossover scale corresponding to $b=8.1$ and $\xi=2/3$.}
\end{figure}

Finally, Fig.~\ref{fig:kolmogorov_equation} illustrates the behavior of the different terms in the normalized Kolmogorov equation (\ref{eq:S3_kolmogorov_regular}) by plotting the third-order structure function along with the inertial and viscous terms, all normalized by $-4/5 \langle \varepsilon \rangle \eta$. For separations larger than the crossover scale, $r/\eta > b$, the third-order structure function matches the inertial term, while the viscous contribution becomes negligible. On the other hand, at small scales ($r/\eta < b$), the viscous term balances the inertial contribution, causing $S_3$ to deviate from the linear scaling and progressively vanish as $r\to 0$.

To verify whether our experimental data follow this phenomenology, we directly compute the mean viscous term, $6\langle \nu \rangle \partial S_2 / \partial r$, from our second-order structure function data. Plotting this quantity together with the third-order structure function, both normalized by $-4/5 \langle \varepsilon \rangle \eta$, shows a good agreement with the theoretical predictions.

This agreement suggests that the modified Kolmogorov equation can capture the behavior of our flows of Carbopol solutions, and that the viscosity fluctuation term $N(r)$ plays a minor role in this balance for the evaluated conditions. Moreover, the agreement of the experimental data with Eq.~(\ref{eq:S3_kolmogorov_normalized}) shows that, as also obtained for $S_2$, the third-order structure function can also be expressed in a self-similar form if the mean viscosity and dissipation rate are considered as those determining the statistics of small scales.

\section{\label{sec:discussion}Discussion}
The parameter $b$ of the Batchelor parametrization, denoting the crossover scale between the dissipative and inertial ranges, was found to be $b\approx 8$ in the previous section from fits to the second- and third-order structure functions. We can compare this value with that corresponding to Newtonian fluids. The Newtonian crossover scale can be obtained by first considering the behavior of the Batchelor parametrization (Eq.~\ref{eq:batchelor_selfsimilar}) for inertial-range separations ($r\gg \eta$) which, adopting $\xi=2/3$, reduces to $S_2/u_\eta^2 = (b^{4/3}/15) (r/\eta)^{2/3}$. Then, equating this result to the Kolmogorov prediction \citep{kolmogorov1991local}, $S_2/u_\eta^2 = C_2 (r/\eta)^{2/3}$, results in $b_{\text{Newtonian}} = (15 C_2)^{3/4}$, where $C_2$ is the Kolmogorov constant. Adopting $C_2 = 2.12$ \citep{sreenivasan_universality_1995} leads to $b_{\text{Newtonian}}\approx 13.4$, which is larger than the value obtained for our Carbopol fluids.

We consider that this difference may be due to two factors. First, the range of separations accessible in this work is limited to $r_{\text{max}}/\eta \approx 13$, which does not cover inertial range separations. Consequently, the fit is dominated by data in the dissipative range, and the resulting parameter estimate may be biased. Second, our measurements correspond to flows at low Reynolds numbers, with a maximum Taylor Reynolds number of approximately 20 (see Table~\ref{tab:table1}), an aspect we discuss further below.

We note that, as originally proposed, the similarity hypotheses are underpinned by the assumption of very large Reynolds numbers \citep{kolmogorov1991local}. However, Kolmogorov phenomenology is also known to successfully describe results corresponding to low Reynolds numbers, with the $-5/3$ spectrum emerging at Taylor-Reynolds numbers as low as $Re_\lambda = 28$ \citep{mccomb2014homogeneous}. \citet{antonia2014collapse} addressed this point, showing that the high $Re_\lambda$ requirement can be significantly relaxed for the first similarity hypothesis to hold. Decreasing the Reynolds number is known to progressively decrease the range of flow scales, shrinking the inertial subrange, and eventually reaching the dissipation range. Considering this, the authors proposed that the first similarity hypothesis can be considered to remain valid as long as the Kolmogorov scaling in the dissipation range applies over a few $\eta$. They subsequently provide a quantitative criterion to determine the required Reynolds number for Kolmogorov scaling to hold for scales smaller than $r/\eta = \mathcal{N}$. Inverting their expression yields $\mathcal{N}_t$, the threshold scale below which the Kolmogorov similarity is expected to hold for a given $Re_\lambda$,

\begin{align}
    \mathcal{N}_t = \left( \frac{4}{3} \sqrt{15} Re_\lambda \right)^{1/2} \label{eq:antonia_FRN}
\end{align}

Evaluating $\mathcal{N}_t$ allows us to assess whether any measured separations in our experiments are influenced by low Reynolds number effects. Because Kolmogorov similarity applies for separations $r/\eta < \mathcal{N}_t$, datasets covering a range beyond this limit ($r_{\text{max}}/\eta > \mathcal{N}_t$) are expected to show departures from the theoretical scaling in the range $\mathcal{N}_t < r/\eta < r_{\text{max}}/\eta$. Conversely, if the maximum measured separation lies below $\mathcal{N}_t$ ($r_{\text{max}}/\eta < \mathcal{N}_t$), the entire measured range remains free from low Reynolds number effects and should follow the theoretical predictions. Considering the Reynolds number of each experimental condition (Table~\ref{tab:table1}), we estimate the corresponding $\mathcal{N}_t$ through Eq.~(\ref{eq:antonia_FRN}). Comparing the obtained $\mathcal{N}_t$ values against the maximum measured separations $r_{\text{max}}/\eta$ shows that low Reynolds number effects are expected only in our three highest-$Re_\lambda$ cases ($Re_\lambda = 16.9, 12.5, \text{and } 9.0$). For these cases, $\mathcal{N}_t \approx 9.4, 8.0, \text{and } 6.8$, whereas the maximum measured separations reach $r_{\text{max}}/\eta \approx 12.9, 9.3, \text{and } 6.8$, respectively. Indeed, for these three cases, the data in the range $\mathcal{N}_t \lesssim r/\eta \lesssim r_{\text{max}}/\eta$ show departures from the analytical predictions. This can be observed more clearly for the third-order structure function in Fig.~\ref{fig:S3}(c), where the experimental data for the three highest-$Re_\lambda$ cases (blue, orange, and yellow filled symbols) fall slightly below both the analytical curve and the DNS data of \citet{rosti2025effect} for $6 \lesssim r/\eta \lesssim 13$, which corresponds to the range where the crossover scale is expected to lie. In contrast, all cases with lower $Re_\lambda$ ($Re_\lambda \le 6.4$) satisfy $r_{\text{max}}/\eta < \mathcal{N}_t$, and are consequently unaffected by low $Re_\lambda$ effects, collapsing onto the analytical predictions, as observed for $S_2/u_\eta^2$ and $S_3/u_\eta^3$.

We consider that the slight departures observed around the crossover scale, likely caused by low Reynolds number effects, combined with the lack of experimental data points within the inertial subrange, may bias the fit of the crossover scale $b$, explaining why our result for Carbopol solutions ($b \approx 8$) is lower than the Newtonian value $b_{\text{Newtonian}} \approx 13.4$.

\section{\label{sec:conclusions}Conclusions}
In this work we experimentally investigated the applicability of Kolmogorov's first similarity hypothesis to shear-dependent viscosity fluids. We analyzed nearly homogeneous and isotropic flows of aqueous solutions of Carbopol, a well-known test fluid exhibiting shear-thinning and yield stress effects, and focused on the behavior of the dissipative scales, where the non-Newtonian viscosity is expected to play a major role. To the best of our knowledge, this study provides the first experimental assessment of the recent extension of the similarity hypotheses proposed by \citet{rosti2025effect}, by evaluating whether the mean viscosity $\langle \nu \rangle$ and mean dissipation rate $\langle \varepsilon \rangle$ serve as universal scaling parameters for small-scale statistics of these complex fluids.

We estimated the mean viscosity $\langle \nu \rangle$ and mean dissipation rate $\langle \varepsilon \rangle$ by combining velocimetry and rheometric data, from which the corresponding Kolmogorov microscales were obtained. Normalizing the second- and third-order structure functions, $S_2$ and $S_3$, by these scales allowed us to collapse the curves corresponding to nine experimental conditions across scales spanning more than two logarithmic decades. This result implies that $S_2$ and $S_3$ can be expressed in a self-similar form when the appropriate parameters, $\langle \nu \rangle$ and $\langle \varepsilon \rangle$, are employed.

We additionally assessed whether analytical expressions describing the self-similar scaling functions for $S_2$ and $S_3$ can be obtained. For the second-order structure function, the Batchelor parametrization \citep{batchelor1951pressure}, an interpolation formula extensively used in Newtonian turbulence \citep{mccomb2014homogeneous}, satisfactorily captures our variable-viscosity results. For the third-order structure function, we evaluated the applicability of a modified Kolmogorov equation incorporating variable viscosity effects. Assuming spatial homogeneity of the mean viscosity, a negligible contribution from the fluctuating viscosity term, and adopting the Batchelor expression for $S_2$ yield an analytical expression showing a very good agreement with our measured $S_3$ curves. We also plotted curves extracted from the DNS results of \citet{rosti2018turbulent} for two different flow indices at higher Reynolds numbers, and observed that these numerical data are in very good agreement with the analytical prediction and our experimental data. Despite the satisfactory performance of the theoretical expressions, fitting the crossover scale $b$ between dissipative and inertial ranges leads to a value $b \approx 8$, which is lower than the standard Newtonian result $b_{\text{Newtonian}} \approx 13.4$. We attribute this discrepancy to low Reynolds number effects and the limited extent of our data into the inertial subrange.

Overall, our results provide the first empirical confirmation of Rosti's \citep{rosti2025effect} extension of the first similarity hypothesis to shear-dependent viscosity fluids. We note that, although the author additionally proposed that small scale statistics depend on the flow index $n$ alongside $\langle \varepsilon \rangle$ and $\langle \nu \rangle$, this dependence is not expected to appear in our second- and third-order structure functions. In the dissipation range, while the dependence on $n$ becomes evident for higher-order statistics (fourth order and above, as shown in Figs.~5(b) and (c) of Ref.~\citep{rosti2025effect}), $S_2$ and $S_3$ remain insensitive to the flow index, as evidenced by the agreement between our experimental results and the numerical data of \citet{rosti2025effect} (Fig.~\ref{fig:S3}(c)).

Finally, we would like to emphasize the importance of experimental investigations of the turbulent behavior of shear-dependent viscosity fluids. While numerical simulations offer powerful capabilities, such as tuning individual rheological parameters (a challenging task to achieve experimentally, where changing fluid concentration simultaneously modifies multiple properties), their validity relies on constitutive models that are mostly not derived from first principles and often oversimplify the complexity of real non-Newtonian fluids. Detailed experimental studies are therefore essential to complement numerical results, with the objective of uncovering the governing physical mechanisms, guiding the development of turbulence models, and refining constitutive relations. In our case, future work will focus on increasing the forcing capabilities of the experimental device, in order to reach higher Reynolds numbers and study inertial-range statistics for these fluids.

\begin{acknowledgments}
The authors thank the Rheology Group of PUC-Rio for their support on the rheological characterization of the test fluids. This work has been supported by ANID/CONICYT FONDECYT Regular 1210936 and ANID/CONICYT FONDEQUIP EQM180150. L.H. acknowledges the scholarship ANIDPFCHA/Doctorado Nacional/2019-21191759.
\end{acknowledgments}

\bibliography{bibfile}

\end{document}